\documentclass{article}
\usepackage{spconf,amsmath,graphicx}
\usepackage{microtype} 
\usepackage{mathpazo} 
\usepackage{caption}
\usepackage{subcaption}
\usepackage{url}
\usepackage{enumitem}
\usepackage{longtable}
\usepackage{xtab}

\title{Triple Attention Network architecture for MovieQA}

\name{Ankit Shah*, Tzu-Hsiang Lin, Shijie Wu }
\address{Carnegie Mellon University \\ Pittsburgh, PA, USA}

\begin{document}


%

\maketitle

\begin{abstract}

Movie question answering, or MovieQA is a multimedia related task
wherein one is provided with a video, the subtitle information, a
question and candidate answers for it. The task is to predict the
correct answer for the question using the components of the multimedia
-- namely video/images, audio and text. Traditionally, MovieQA is done
using the image and text component of the multimedia. In this paper,
we propose a novel network with triple-attention architecture for the
inclusion of audio in the Movie QA task. This architecture is
fashioned after a traditional dual attention network focused only on
video and text. Experiments show that the inclusion of audio using the
triple-attention network results provides complementary information for Movie QA
task which is not captured by visual or textual component in the data. Experiments with a wide range of audio features show that using such a network can indeed improve MovieQA
performance by about 7\% relative to just using only visual features.

\end{abstract}


\section{Introduction}
Vision and language have always been an essential part of human intelligence. With the development of more advanced systems, machines have been catching up with humans perception abilities. For example, image classification tasks, LeNet \cite{lecun2015lenet} and ResNet \cite{DBLP:journals/corr/HeZRS15} are on par with human evaluation. However, machines are still far behind human regarding comprehension abilities. Visual question answering \cite{antol2015vqa} is task that tests machine ability to comprehend natural languages. Many approaches has been proposed for this task \cite{DBLP:journals/corr/ZhouTSSF15, DBLP:journals/corr/NohSH15, DBLP:journals/corr/FukuiPYRDR16, DBLP:journals/corr/LuYBP16, DBLP:journals/corr/NamHK16}. Among them, the dual attention model combines the semantics of images and texts and reached the 1st place in the MovieQA challenge\cite{MovieQA}.  Inspired by this model, we propose two three-fold attention architectures to exploit the power of audio. 

The Movie-QA challenge\cite{MovieQA} was proposed for deep machine comprehension of complex scenes.  Unlike Visual Question Answering\cite{agrawal2015vqa} where the model only needs to understand the object in a single image, understanding a movie requires pooling information from multiple video frames and align them with the questions and answers.   Previous works \cite{DBLP:journals/corr/NamHK16, na2017read} focus on attention modules that align certain parts of videos to certain parts of the text.  These works have achieved remarkable results and helped us gain understanding about the interaction between vision and language, but does not include the audio information.

When watching movies, humans not only attend to the scenes and dialogue, but also listen to the sound/music accompanies by the scenes, and the audio include information that are not captured in vision and text.  The sound of a gun shot indicates that the gun has been fired, whereas an image with a gun alone only indicates the presence of the object.  Furthermore, if an event like a car accident happens off-screen, there is no indication from the scene or the subtitles that an accident happened, but we can hear the sound of the vehicles crashing into each other. By adding audio information, we make use of every information we can find in a movie. With this hypothesis, we propose to combine the previously neglected audio information into Movie-QA.  We believe our work is interesting and novel in the following aspects: 1) We are the first to incorporate audio information into Movie-QA. and 2) We introduce a triple attention network that combines heterogeneous information sources including question, subtitles, video and audio. Our work can be seen as combining two fields, 1). Audio Event Detection and 2). Visual Question Answering.  Our work also falls under the broader category of Multi-modal Machine Learning\cite{baltruvsaitis2018multimodal}.


MovieQA is a video question answering challenge that, given video, audio, subtitles and the question, the model has to predict the right output. In the question set with video clips, the training set, validation set, and the test set has 4385, 1098, and 1288 questions. Each question has 5 answers, only one of them is correct. It's been showed that people have only 24.7\% given only the question and answers \cite{MovieQA}.



\section{Methodology}

\subsection{Audio features} \label{ssec:feature}
We used the below three different approaches to extract audio features Mel Spectrograms, SoundNet\cite{aytar2016soundnet} and Weakly Labelled Audio Network WALNet based CNN features \cite{2017arXiv171101369K,2018arXiv180409288S}.

\begin{figure}[h!]
\centering
\includegraphics[width=0.5\textwidth]{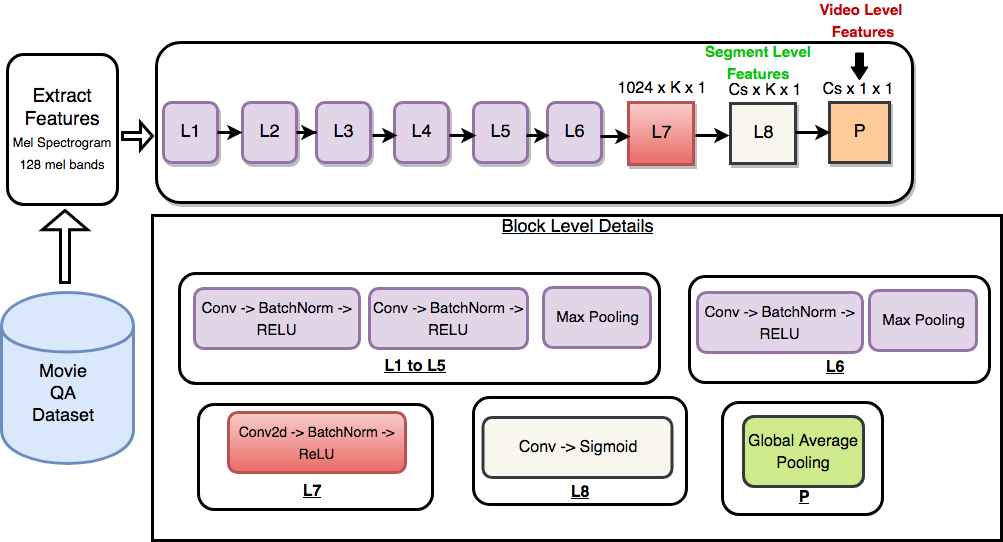} 
\caption{WALNet based audio features extraction for Movie-QA. We extract the layer 8 output which is the output at segment level for the recording}
\label{fig:audiofeatureanalysis}
\end{figure}

Fig [\ref{fig:audiofeatureanalysis}] describes WALNet audio feature extraction. The blocks of layers L1 to L5 consists of two convolutional layers with batch normalization before non linear activation activation followed by a max pooling layer. ReLU \cite{nair2010rectified} ($max(0,x)$) is used in all layers from L1 to L6. $3 \times 3$  convolutional filters are used in all layers from L1 to L6. Stride and padding values are fixed to $1$. The number of filters employed in different layers is as follows, \emph{\{L1: 16, L2:32, L3:64, L4:128, L5:256, L6:512 \}}. The max pooling operation is applied over a window of size $2 \times 2$. Logmel inputs are treated as single channel inputs. The max pooling operation reduces the size by a factor of $2$. For example, a recording consisting of $128$ logmel frames that is an input of size $X \in R^{1 \times 128 \times 128}$, will produce and output of size $16 \times 64 \times 64$ after L1.  This design aspect applies  to all layer blocks from L1 to L6 and hence $1 \times 128 \times 128$ input will produce a $512 \times 2 \times 2$ output after L6. L7 is a convolutional layer with $1024$ filters of size $2 \times 2$. Once again ReLU activation is used. Stride is again fixed to $1$ and no padding is used. 

Layer L8 represents segment level output. It is a convolutional layer consisting of $C_s$ filters of size $1 \times 1$. $C_s$ is the number of class in the dataset. The final \emph{global pooling player}, $P$, maps the segment level outputs from L8 to full recording level outputs. In this case we use \emph{average} function to perform this mapping. Layer L8 represents the segment level output and Layer P represents the video level output. The amount of segment over which we get the output is segment size of 128 frames or 1.5 seconds and the segment can be thought of as moving by 64 frames with a consecutive overlap of 50 $\%$ Hence, the networks is able to localize events despite learning from weak labeled data. We extract the segment level output at layer 8 for the analysis of the feature representation in Movie QA task.

\begin{table}
\centering
\resizebox{1.0\columnwidth}{!}{
\begin{tabular}{ |c|c| } \hline
Feature & Description \\
\hline 
 \hline 
 Mel Spectrogram & Based on human auditory perception \\ 
 \hline 
 SoundNet & Recognize objects and scenes from sound \\ 
 \hline
 WALNet & Based on VGG net for audio event detection \\ 
 \hline
\end{tabular}
}

\end{table}

\subsection{Video features and Textual features}
Videos are processed as a sequence of images and fed into the two-stream ConvNet \cite{DBLP:journals/corr/SimonyanZ14}. While the spatial stream ConvNet directly operates on still images, the temporal stream ConvNet process the images into optical flow first, then feed it into the convolution layer. We first calculate the displacement vector of each point in the frame based on brightness. Then separate them into the horizontal and vertical placement and feed it into the ConvNet. We employ 200 dimension pre-trained GloVe\cite{pennington2014glove} word embeddings trained on 6 billion Wikipedia corpus as initial weights to our neural network. There are more slangs in movie lines, we expect to see it change through training. 
 


\subsection{Dual Attention Networks} 
Dual Attention Network(DAN) was proposed for Visual Question Answering\cite{antol2015vqa} and Image Text Matching. The key idea of a Dual Attention Network\cite{dan} is to produce joint embedding memory vector $m_k$ that contains both visual and textual information.
Suppose we have visual features $\{v_1,v_2,...,v_N\}$ and text features $\{u_1,u_2,...,u_T\}$.  Here $N$ corresponds to number of different image regions and $T$ corresponds to the number of words. The initial memory vector $m_k$ will be first be initialized with the visual features and textual features.  At the next hop $k+1$, $m_k$ will be used to attend to the visual and text inputs.  Using the attention weights, we can compute attention context vectors for visual and textual information.  Then using these attention context vectors, we compute a new memory vector $m_{k+1}$.  The memory update rule can be written as 
\begin{equation} \label{eq:1}
    m_{k+1} = m_k + q * v
\end{equation}
Here $q$,$v$ denotes the textual and visual attention context vectors. The final memory vector $m_K$ will be fed into a  scoring network for question answering. 


\subsection{Triple Attention Network - Dual Is Not Enough}
In the MovieQA dataset, every question is paired with its corresponding video clips, subtitles of these video clips, audio, and also the answer options.  Expanding from the Dual Attention Network, our memory vector will use question, video, subtitles, and audio as features. The initial memory vector will be initialized using these four features, and attention weights will be computed over these four for new context vectors.  For the answers, we encode every answer option using an LSTM and concatenate the encoded answer vector with the final memory vector and feed the concatenated vector into a feed forward neural network to produce the final scores.

Due to the increase number of modalities, we will need a new memory update rule.  We designed the update rule written in equation \ref{eq:2}.  Here q, v, s, a denotes the question, video, subtitle and audio information respectively.  $\lambda$s are parameters we define for experiment analysis.  For example, $\lambda={0,1,1}$ means that we use subtitle and audio information to answer the questions.

\begin{equation} \label{eq:2}
    m_{k+1} = m_k + \lambda_1 q * v + \lambda_2 q * s + \lambda_3 q * a
\end{equation}

\subsubsection{Feature Encoding}
In a video clip, the frames can be long as to 100 video frames.  Also the length of the subtitles can be up to 1k.  Computing attention over every frame and every word in the subtitles will be impractical, since most of the frame overlap, words in subtitles are not as important as words in the question, and may result in GPU memory limitation.  Therefore, after extracting video, audio features and project subtitles into word embeddings, we first pass these raw features into CNNs for length reduction and also learn higher level representations.  During memory computation, the attention will be based on the output of the CNNs.

\begin{figure}[h!]
    \centering
    \includegraphics[width=0.5\textwidth]{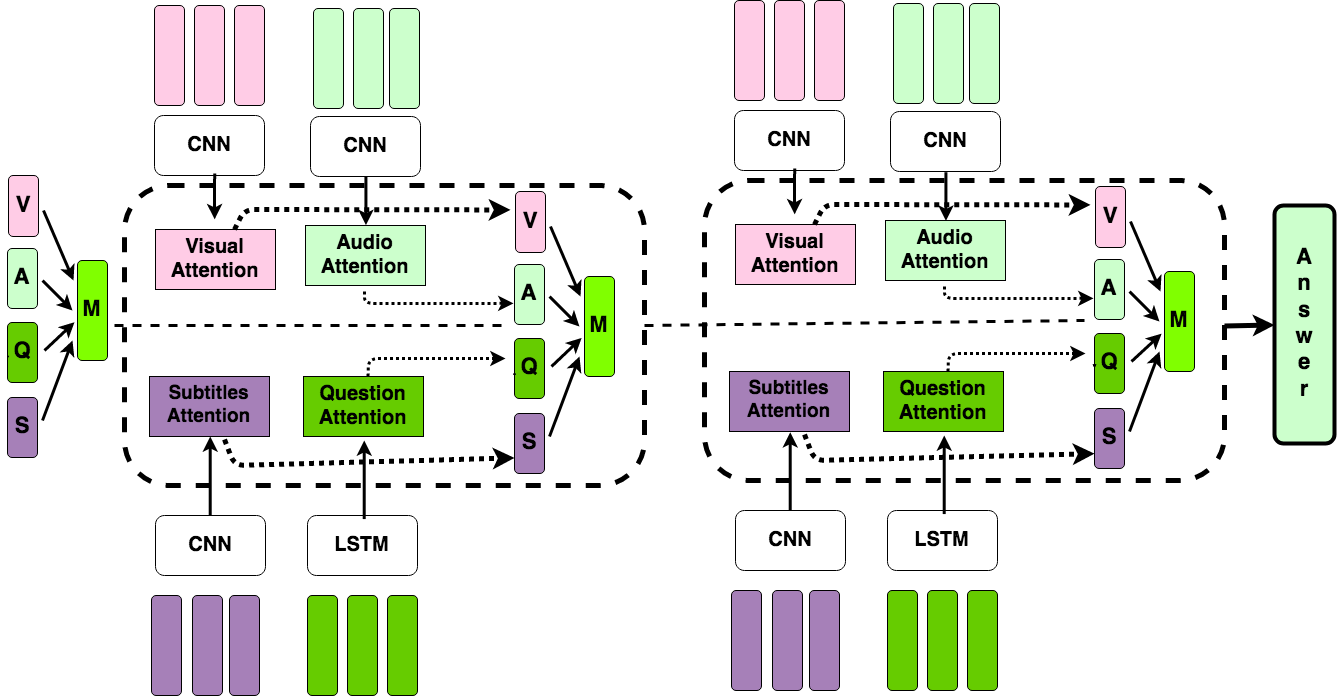} 
    \caption{MovieQA Network architecture using the modalities and their respective attention. Final memory vector is used to predict the correct answer for a given question}
    \label{fig:movieDAN}
\end{figure}

\section{Experimental Setup}

\subsection{MovieQA Dataset}

Here we describe the video+subtitles track of the MovieQA dataset\cite{MovieQA}. In the video-subtitle track of the challenge, there are 4385 and 1085 questions respectively in training and validation set. Each question is incorporated with one or more video clips, and subtitles for the clips. And each question is provided with 5 answers. These answers are highly challenging, human without seeing the movie clips can only achieve $24.7\%$ accuracy \cite{MovieQA}. Out of the 408 movies, 140 of them contain video information, which has 6796 video clips.  However, 303 of the video clips contain no audio information.  We will focus on comparing the movies with audio information.


\subsection{Feature Extraction}

\textbf{Question}: For every question, we first project every word into word embeddings.  The word embeddings will then be fed into a bidirectional LSTM.  The outputs of the LSTM is used to compute attention.
\noindent \textbf{Video}:  We feed each video frame into VGG to obtain their corresponding features.
\noindent \textbf{Subtitles}: Similar to question, we project each word into word embeddings.  Then we pass through a 2 layer CNN.  The output of the CNN is used to compute attention.
\noindent \textbf{Audio}: We format the audio to 16 bit resolution, mono channel audio resampled at 44.1kHz.  We set our audio feature dimension to 128 and extracted the features mentioned in section \ref{ssec:feature}.

\subsection{Dual Attention Network Implementation}

We implemented the Dual Attention Network in PyTorch\cite{paszke2017automatic}.  To verify our implementation, we followed the hyperparameter settings in \cite{dan} and tested on the VQA\cite{agrawal2015vqa} v1 dataset.  Our implementation can be found on Github\footnote{https://github.com/iammrhelo/pytorch-vqa-dan}. 

\subsection{Hyperparameter Details} 

We set all word embedding size to 128 and hidden size to 32.  We shared word embedding for all question, subtitle and answer encoders.  We set dropout rate to 0.5.  We used Adam with learning rate 0.001 and trained for 50 epochs and cross-entropy loss after the last softmax layer. During training, we randomly shuffle the five options for every question. 

\section{Results and Discussion}




\subsection{Multimodal Fusion Comparison}
The accuracy is computed as the ratio of correct answers out of all answers. Since there are 5 options for every question, a random agent would have about 20\% accuracy. The current state of the art accuracy is 47.6\% \cite{liu2020dual} on the test set. We tested on different combinations of modalities.  In equation \ref{eq:2}, we can see that the subtitles, video, and audio all attend to the questions.  For $Q+V, Q+S, Q+A$, there is only one $\lambda$ that is nonzero and set to 1.  For $Q+V+S$, $Q+S+A$, there are 2 $lambda$s that are non zero.  For $Q+V+S+A$, all 3 $\lambda$s are used. Among all the single modality models($Q+V$,$Q+S$,$Q+A$), \textbf{Q + A} achieves the best accuracy. This show that audio provide a rich source of information in MovieQA.


While the single modality models has good accuracy, the results of our multiple modality models don't have accuracy gains. The first hypothesis is that our memory update rule \ref{eq:2} is not very effective. Another hypothesis is that the level of abstraction of our input data are different, and directly multiplying them together will not be effective.  For example, the question input focuses on word level, whereas in the subtitles, a CNN is pooled across multiple words and sentences to create a higher level representation.  Also, the results of our video features do not work well. In Movie-QA, most questions are related to the plot, and simply knowing what objects appear in the video is not sufficient to understand the actions of a certain character.


\begin{table}[h]

\begin{center}
\resizebox{1.0\columnwidth}{!}{
\begin{tabular}{ |c|c|c|c| }
 \hline 
 Accuracy & Videos & Subtitles & Audios \\ 
 \hline 
 Random & \multicolumn{3}{|c|}{20.00} \\ 
 \hline 
 MovieQA Leaderboard 1st place & \multicolumn{3}{|c|}{47.61} \\ 
 \hline
 Q + V &   34.7 & - & -\\ 
 \hline
 Q + S &   - & 35.90 & -\\ 
 \hline 
 Q + A &  -  & - & 37.2 \\ 
 \hline
 Q + V + S & \multicolumn{2}{|c|}{34.7} & - \\ 
 \hline
 Q + S + A & - &  \multicolumn{2}{|c|}{34.9} \\ 
 \hline
 Q + V + S + A & \multicolumn{3}{|c|}{37.60} \\
 \hline
\end{tabular}
\vspace{2mm}
\label{tab:tab_1}

}
\end{center}
\caption{Accuracies of Different Combinations of Input Data Source}
\end{table}

\subsection{Audio Feature Comparison}

For the Q + A model, we compared the accuracies of different audio input. In the first table, different layers of SoundNet network is extracted and fed into the model. We show that the 24th layer provide the best sound features. In the second feature, we compared the accuracies of MelSpectrogram, SoundNet and WALNet features. And we prove that WALNet feature is the best among those three.   Since all the audio features we extracted has around 30\% accuracy, we can conclude that audio event detection, no matter the form, is beneficial to question answering that contains audio information.

\begin{center}
\begin{tabular}{ |c|c|c|c|c| } 
 \hline 
    & Q+SN-14 & Q+SN-15 & Q+SN-16 & Q+SN-24\\ 
 \hline 
 Accuracy &  34.62 & 26.45 & 35.05 & 36.6\\ 
 \hline
\end{tabular}
\\\vspace{2mm}Table 2: Accuracy of Different Layers of SoundNet Features
\label{tab:tab_2}
\end{center}

\begin{center}
\begin{tabular}{ |c|c|c|c| } 
 \hline 
    & Q+MelSpec & Q+SN-24 & Q+WALNet\\ 
 \hline 
 Accuracy &  31.83 & 36.6 & 37.2 \\ 
 \hline
\end{tabular}
\\\vspace{2mm}Table 3: Accuracy of Different Types of Audio Input Features
\label{tab:tab_3}
\end{center}

\subsection{Detail Analysis: Subtitles v.s. Audio}
A closer inspection on the predictions of different input models shows that subtitle based model (Q + S) has a higher accuracy over factual based questions and audio based model (Q + A) has a higher accuracy over action based questions. They are complimentary in the whole question set. 

In the first example, the way how batman survives Joker's bullets is captured in a motion scene filled with sounds like gun shots. The movie scene doesn't have accompanied subtitles which explains batman's behavior. 

\textbf{Example 1}. Q + A model right but Q + S model wrong\\
\textbf{Movie}: Batman (1989)\\
\textbf{Question}: How does Bruce survive the Joker's bullets?\\
\textbf{Answers}: "He runs really fast and dodges the bullets", "The Joker misses him because he has really bad aim", "He is wearing body armor", "He is impervious to bullets", "He uses Batarangs to deflect the bullets" = Action/Activity based right for audio. 

\textbf{Example 2} Q + S model wrong and Q + A model right \\
\textbf{Question}: What happens to the father in a ruined town? \\
\textbf{Answers}: 'He is shot in the leg with an arrow', 'He faints', 'He is attacked with a flare gun', 'He is robbed', 'He is shot with a gun'\\



\section{Conclusion}
In this paper, we show that audio data is a rich source of information and should be utilized in Movie-QA challenge. We propose a novel way to fuse multimodal attention features and audio provides complementary information in comparison with. Our exploration will lead to better architectures in the community for MovieQA and take cues from our work to advance state of the art. 

\ninept
\bibliographystyle{IEEE}
\bibliography{refs}

\begin{thebibliography}{10}

\bibitem{lecun2015lenet}
Yann LeCun et~al.,
\newblock ``Lenet-5, convolutional neural networks,''
\newblock .

\bibitem{DBLP:journals/corr/HeZRS15}
Kaiming He, Xiangyu Zhang, Shaoqing Ren, and Jian Sun,
\newblock ``Deep residual learning for image recognition,''
\newblock {\em CoRR}, vol. abs/1512.03385, 2015.

\bibitem{antol2015vqa}
Stanislaw Antol, Aishwarya Agrawal, Jiasen Lu, Margaret Mitchell, Dhruv Batra,
  C~Lawrence~Zitnick, and Devi Parikh,
\newblock ``Vqa: Visual question answering,''
\newblock in {\em Proceedings of the IEEE International Conference on Computer
  Vision}, 2015, pp. 2425--2433.

\bibitem{DBLP:journals/corr/ZhouTSSF15}
Bolei Zhou, Yuandong Tian, Sainbayar Sukhbaatar, Arthur Szlam, and Rob Fergus,
\newblock ``Simple baseline for visual question answering,''
\newblock {\em CoRR}, vol. abs/1512.02167, 2015.

\bibitem{DBLP:journals/corr/NohSH15}
Hyeonwoo Noh, Paul~Hongsuck Seo, and Bohyung Han,
\newblock ``Image question answering using convolutional neural network with
  dynamic parameter prediction,''
\newblock {\em CoRR}, vol. abs/1511.05756, 2015.

\bibitem{DBLP:journals/corr/FukuiPYRDR16}
Akira Fukui, Dong~Huk Park, Daylen Yang, Anna Rohrbach, Trevor Darrell, and
  Marcus Rohrbach,
\newblock ``Multimodal compact bilinear pooling for visual question answering
  and visual grounding,''
\newblock {\em CoRR}, vol. abs/1606.01847, 2016.

\bibitem{DBLP:journals/corr/LuYBP16}
Jiasen Lu, Jianwei Yang, Dhruv Batra, and Devi Parikh,
\newblock ``Hierarchical question-image co-attention for visual question
  answering,''
\newblock {\em CoRR}, vol. abs/1606.00061, 2016.

\bibitem{DBLP:journals/corr/NamHK16}
Hyeonseob Nam, Jung{-}Woo Ha, and Jeonghee Kim,
\newblock ``Dual attention networks for multimodal reasoning and matching,''
\newblock {\em CoRR}, vol. abs/1611.00471, 2016.

\bibitem{MovieQA}
Makarand Tapaswi, Yukun Zhu, Rainer Stiefelhagen, Antonio Torralba, Raquel
  Urtasun, and Sanja Fidler,
\newblock ``{MovieQA: Understanding Stories in Movies through
  Question-Answering},''
\newblock in {\em IEEE Conference on Computer Vision and Pattern Recognition
  (CVPR)}, 2016.

\bibitem{agrawal2015vqa}
Aishwarya Agrawal, Jiasen Lu, Stanislaw Antol, Margaret Mitchell, C~Lawrence
  Zitnick, Dhruv Batra, and Devi Parikh,
\newblock ``Vqa: Visual question answering,''
\newblock {\em arXiv preprint arXiv:1505.00468}, 2015.

\bibitem{na2017read}
Seil Na, Sangho Lee, Jisung Kim, and Gunhee Kim,
\newblock ``A read-write memory network for movie story understanding,''
\newblock {\em arXiv preprint arXiv:1709.09345}, 2017.

\bibitem{baltruvsaitis2018multimodal}
Tadas Baltru{\v{s}}aitis, Chaitanya Ahuja, and Louis-Philippe Morency,
\newblock ``Multimodal machine learning: A survey and taxonomy,''
\newblock {\em IEEE Transactions on Pattern Analysis and Machine Intelligence},
  2018.

\bibitem{aytar2016soundnet}
Yusuf Aytar, Carl Vondrick, and Antonio Torralba,
\newblock ``Soundnet: Learning sound representations from unlabeled video,''
\newblock in {\em Advances in Neural Information Processing Systems}, 2016.

\bibitem{2017arXiv171101369K}
A.~{Kumar}, M.~{Khadkevich}, and C.~{Fugen},
\newblock ``{Knowledge Transfer from Weakly Labeled Audio using Convolutional
  Neural Network for Sound Events and Scenes},''
\newblock {\em ArXiv e-prints}, Nov. 2017.

\bibitem{2018arXiv180409288S}
A.~{Shah}, A.~{Kumar}, A.~G. {Hauptmann}, and B.~{Raj},
\newblock ``{A Closer Look at Weak Label Learning for Audio Events},''
\newblock {\em ArXiv e-prints}, Apr. 2018.

\bibitem{nair2010rectified}
Vinod Nair and Geoffrey~E Hinton,
\newblock ``Rectified linear units improve restricted boltzmann machines,''
\newblock in {\em Icml}, 2010.

\bibitem{DBLP:journals/corr/SimonyanZ14}
Karen Simonyan and Andrew Zisserman,
\newblock ``Two-stream convolutional networks for action recognition in
  videos,''
\newblock {\em CoRR}, vol. abs/1406.2199, 2014.

\bibitem{pennington2014glove}
Jeffrey Pennington, Richard Socher, and Christopher Manning,
\newblock ``Glove: Global vectors for word representation,''
\newblock in {\em Proceedings of the 2014 conference on empirical methods in
  natural language processing (EMNLP)}, 2014, pp. 1532--1543.

\bibitem{dan}
Hyeonseob Nam, Jung{-}Woo Ha, and Jeonghee Kim,
\newblock ``Dual attention networks for multimodal reasoning and matching,''
\newblock {\em CoRR}, vol. abs/1611.00471, 2016.

\bibitem{paszke2017automatic}
Adam Paszke, Sam Gross, Soumith Chintala, Gregory Chanan, Edward Yang, Zachary
  DeVito, Zeming Lin, Alban Desmaison, Luca Antiga, and Adam Lerer,
\newblock ``Automatic differentiation in pytorch,''
\newblock 2017.

\bibitem{liu2020dual}
Fei Liu, Jing Liu, Xinxin Zhu, Richang Hong, and Hanqing Lu,
\newblock ``Dual hierarchical temporal convolutional network with qa-aware
  dynamic normalization for video story question answering,''
\newblock in {\em Proceedings of the 28th ACM International Conference on
  Multimedia}, 2020, pp. 4253--4261.

\end{thebibliography}

\end{document}